\documentclass[conference]{IEEEtran}

\usepackage[cmex10]{amsmath}
\usepackage{amsthm,amssymb}
\usepackage[pdftex]{graphicx}
\usepackage{color}
\usepackage{cite}

\title{Strong Secrecy for Erasure Wiretap Channels} 

\author{\IEEEauthorblockN{Ananda T. Suresh\IEEEauthorrefmark{1},
Arunkumar Subramanian\IEEEauthorrefmark{2},
Andrew Thangaraj\IEEEauthorrefmark{1}, 
Matthieu Bloch\IEEEauthorrefmark{2} and
Steven McLaughlin\IEEEauthorrefmark{2}}
\IEEEauthorblockA{\IEEEauthorrefmark{1}	Department of Electrical Engineering, Indian Institute of Technology, Madras \\
																				Email: andrew@iitm.ac.in
																			}
\IEEEauthorblockA{\IEEEauthorrefmark{2}	School of Electrical and Computer Engineering, Georgia Institute of Technology, USA and GT-CNRS UMI 2958, France\\
																				Email: arunkumar@gatech.edu, matthieu.bloch@ece.gatech.edu, swm@ece.gatech.edu
																			}
}
\DeclareMathAlphabet{\eurm}{U}{eur}{m}{n}
\DeclareMathAlphabet{\mathbsf}{OT1}{cmss}{bx}{n}
\DeclareMathAlphabet{\mathssf}{OT1}{cmss}{m}{sl}
\DeclareMathAlphabet{\mathcsf}{OT1}{cmss}{sbc}{n}

\newcommand{\samplevalue}[1]{\eurm{\lowercase{#1}}}
\newcommand{\randomvalue}[1]{\eurm{\uppercase{#1}}}

\DeclareSymbolFont{bsfletters}{OT1}{cmss}{bx}{n}  
\DeclareSymbolFont{ssfletters}{OT1}{cmss}{m}{n}
\DeclareMathSymbol{\bsfGamma}{0}{bsfletters}{'000}
\DeclareMathSymbol{\ssfGamma}{0}{ssfletters}{'000}
\DeclareMathSymbol{\bsfDelta}{0}{bsfletters}{'001}
\DeclareMathSymbol{\ssfDelta}{0}{ssfletters}{'001}
\DeclareMathSymbol{\bsfTheta}{0}{bsfletters}{'002}
\DeclareMathSymbol{\ssfTheta}{0}{ssfletters}{'002}
\DeclareMathSymbol{\bsfLambda}{0}{bsfletters}{'003}
\DeclareMathSymbol{\ssfLambda}{0}{ssfletters}{'003}
\DeclareMathSymbol{\bsfXi}{0}{bsfletters}{'004}
\DeclareMathSymbol{\ssfXi}{0}{ssfletters}{'004}
\DeclareMathSymbol{\bsfPi}{0}{bsfletters}{'005}
\DeclareMathSymbol{\ssfPi}{0}{ssfletters}{'005}
\DeclareMathSymbol{\bsfSigma}{0}{bsfletters}{'006}
\DeclareMathSymbol{\ssfSigma}{0}{ssfletters}{'006}
\DeclareMathSymbol{\bsfUpsilon}{0}{bsfletters}{'007}
\DeclareMathSymbol{\ssfUpsilon}{0}{ssfletters}{'007}
\DeclareMathSymbol{\bsfPhi}{0}{bsfletters}{'010}
\DeclareMathSymbol{\ssfPhi}{0}{ssfletters}{'010}
\DeclareMathSymbol{\bsfPsi}{0}{bsfletters}{'011}
\DeclareMathSymbol{\ssfPsi}{0}{ssfletters}{'011}
\DeclareMathSymbol{\bsfOmega}{0}{bsfletters}{'012}
\DeclareMathSymbol{\ssfOmega}{0}{ssfletters}{'012}

\newcommand{\rvG}{{\randomvalue{G}}}

\newcommand{\rvK}{{\randomvalue{K}}}
\newcommand{\rvM}{{\randomvalue{M}}}	
\newcommand{\rvX}{{\randomvalue{X}}}  	
\newcommand{\rvZ}{{\randomvalue{Z}}}	

\newcommand{\svs}{{\samplevalue{s}}}

\newcommand{\rvx}{{\randomvalue{x}}}	
\newcommand{\svx}{{\samplevalue{x}}}

\newcommand{\svz}{{\samplevalue{z}}}

\newcommand{\calG}{{\mathcal{G}}}

\newcommand{\calO}{{\mathcal{O}}}

\newcommand{\calS}{{\mathcal{S}}}

\renewcommand{\P}[2][]{{\mathbb{P}_{#1}}{\left[#2\right]}}

\newcommand{\avgI}[1]{{{\mathbb{I}}\!\left(#1\right)}}

\newcommand{\avgH}[1]{{\mathbb{H}}\!\left(#1\right)}

\newtheorem{lemma}{Lemma}

\newtheorem{theorem}{Theorem}

\newtheorem{remark}{Remark}
\newtheorem{corollary}{Corollary}

\begin{document}
\maketitle

\begin{abstract}
We show that duals of certain low-density parity-check (LDPC) codes, when used in a standard coset coding scheme, provide strong secrecy over the binary erasure wiretap channel (BEWC). This result hinges on a stopping set analysis of ensembles of LDPC codes with block length $n$ and girth $\geq 2k$, for some $k \geq 2$. We show that if the minimum left degree of the ensemble is $l_\mathrm{min}$, the expected probability of block error is $\calO(\frac{1}{n^{\lceil l_\mathrm{min} k /2 \rceil - k}})$ when the erasure probability $\epsilon < \epsilon_\mathrm{ef}$, where $\epsilon_\mathrm{ef}$ depends on the degree distribution of the ensemble. As long as $l_\mathrm{min} > 2$ and $k > 2$, the dual of this LDPC code provides strong secrecy over a BEWC of erasure probability greater than $1 - \epsilon_\mathrm{ef}$.
\end{abstract}

\section{Introduction}
\label{sec:introduction}
The information-theoretic limits of secure communications over public channels were first investigated by Shannon~\cite{Shannon1948a}; given a message $\rvM$ and its corresponding cryptogram $\rvX^n$ of length $n$, a message is communicated with \emph{perfect secrecy} if $\avgI{\rvM;\rvX^n}=0$. Shannon proved the disappointing result that perfect secrecy requires a secret key $\rvK$ with entropy $\avgH{\rvK}\geq\avgH{\rvM}$. In this setting, Wyner subsequently proposed an alternative model for secure communication called a \emph{wiretap channel}~\cite{Wyner1975}, in which all communications occur over noisy channels and the eavesdropper observes a degraded version $\rvZ^n$ of the signal received by the legitimate receiver. Wyner introduced the notion of \emph{weak secrecy}, which requires the leaked information \emph{rate} $\frac{1}{n}\avgI{\rvM;\rvZ^n}$ to vanish as $n\rightarrow\infty$, and established the \emph{weak secrecy capacity}, that is the maximum secure communication rate achievable over a wiretap channel under this condition. Maurer and Wolf later highlighted the shortcomings of weak secrecy for cryptographic purposes, and suggested to replace it with the notion of \emph{strong secrecy}, by which the absolute information $\avgI{\rvM;\rvZ^n}$ should vanish as $n\rightarrow\infty$. Surprisingly, this stronger secrecy requirement does not reduce secrecy capacity~\cite{Maurer2000,Csiszar1996}.

Despite the surge of recent results investigating wiretap channels, the design of coding schemes with provable secrecy rate has not attracted much attention. Some efforts in coding for wiretap channels include \cite{Bennett1995, Ozarow1984, Thangaraj2007, Liu2007, Cohen2006}.

In this work, we revisit the LDPC-based coset coding scheme of~\cite{Thangaraj2007} for the binary erasure wiretap channel. We first show that the dual of randomly generated LDPC codes can achieve strong secrecy provided the probability of block error of the LDPC codes decays faster than $\frac{1}{n}$ with the block length $n$ in a binary erasure channel. Then, we show that for certain small-cycle-free LDPC ensembles, the probability of block error under iterative decoding decays as $\calO(\frac{1}{n^2})$. We obtain this result by analyzing the stopping sets of LDPC ensembles. Stopping sets \cite{Di2002, Richardson2001c} determine whether iterative decoding of LDPC codes under erasures will succeed or not. Asymptotic enumeration of stopping sets has been done by several authors (see \cite{Orlitsky2005,Olgica,Burshtein,ModernCodingTheory} and references thereof). We follow the approach in \cite{Orlitsky2005}, where asymptotics of the average block error probability of LDPC codes were derived.

Ensembles of LDPC codes with better than $\frac{1}{n}$ average block error probability are known from prior studies which use expander-based ideas and stopping set expurgation \cite{Korada,Amraoui2009}. Expander-based ideas typically require minimum bit node degree of five or above resulting in a decrease in thresholds. Expurgation of stopping sets is usually more difficult to achieve than expurgation of short cycles in random constructions. In our approach, we consider ensembles with finite girth. Restricting the girth results in $\calO(\frac{1}{n^2})$ expected block error probability in irregular ensembles with minimum girth 4 and minimum bit node degree 3. This enables high erasure thresholds and efficient construction methods.  

In this work, the code construction for strong secrecy is fundamentally different from Maurer and Wolf's procedure to obtain strong secrecy from weak secrecy~\cite{Maurer2000}. Maurer and Wolf's method relies on the equivalence of key-generation with one-way communication and coding for the wiretap channel, while our code construction yields a forward error-control scheme directly. Nevertheless, the constraint imposed in our code construction limits the achievable secrecy rate.

The rest of the paper is organized as follows. In Section~\ref{sec:coding-secrecy}, we briefly review the coset coding scheme for the binary erasure wiretap channel and establish the connection between strong secrecy and the decay of probability of block error with code length. In Section~\ref{sec:ldpc-nosmallcycles}, we show that the probability of block error for ensembles without short cycles decays fast enough to guarantee strong secrecy.

\section{Secrecy Coding for the Binary Erasure Wiretap Channel}
\label{sec:coding-secrecy}

\begin{figure}[ht]
  \centering
  \scalebox{0.65}{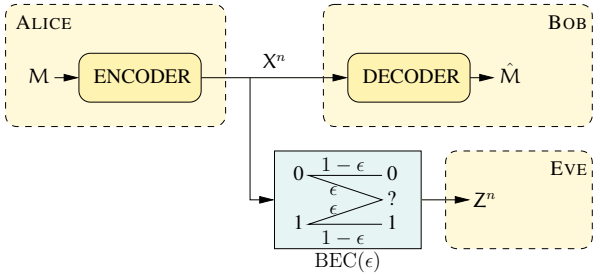}
  \caption{Binary erasure wiretap channel.}
  \label{fig:bewtc}
\end{figure}

The wiretap channel considered in this work, denoted by $\mathrm{BEWC}(\epsilon)$, is illustrated in Fig.~\ref{fig:bewtc}. The channel between the legitimate parties is noiseless while the eavesdropper's channel is a binary erasure channel with erasure probability $\epsilon$ (denoted BEC$(\epsilon)$). The secrecy capacity of this wiretap channel is $C_s = \epsilon$~\cite{Wyner1975}.

The ``coset coding'' scheme to communicate secretly over this channel, proposed in~\cite{Ozarow1984}, is the following. Prior to transmission, Alice and Bob agree on a $(n,n-k)$ code $C$ with parity check matrix $\mathbf{H}$. The coset of $C$ with syndrome $\svs^k$ is denoted by $C(\svs^k)=\{\svx^n\in\{0,1\}^n:\svs^k=\mathbf{H}^T\svx^n\}$. To transmit a message $\rvM$ of $k$ bits, Alice transmits a codeword $\rvX^n$ chosen uniformly at random in $C(\rvM)$. Bob decodes his received codeword $\rvX^n$ by forming the syndrome $\mathbf{H}^T\rvx^n$.

The following theorem due to Ozarow and Wyner connects the equivocation of the eavesdropper to algebraic properties of the generator matrix.
\begin{theorem}[\cite{Ozarow1984}]
\label{th:ozarow}
Let $C$ be a $(n,n-k)$ code with generator matrix $\mathbf{G}=\left[g_1,\dots,g_n\right]$, where $g_i$ represents the $i$-th column of $\mathbf{G}$. Let $\svz^n$ be an observation of the eavesdropper with $\mu$ unerased position given by $\{i:\svz_i\neq ?\}=\{i_1,\dots,i_\mu\}$. Let $\mathbf{G}_\mu=[g_{i_1}\dots g_{i_\mu}]$. Then, $\avgH{\rvM|\svz^n}=k$ iff $\mathbf{G}_\mu$ has full rank.
\end{theorem}

Based on Theorem~\ref{th:ozarow}, we can now connect the rate of convergence of $\avgI{\rvM;\rvZ^n}$ to the probability that a submatrix of $\mathbf{G}$ has full rank.
\begin{lemma}
\label{lm:secrecy_perr_connection}
  Let $\rvG_\mu$ be the submatrix of $\mathbf{G}$ corresponding to the unerased positions in $\rvZ^n$. Let $p_{nf}$ be the probability that $\rvG_\mu$ is not full rank. Then, a coset coding scheme operates with strong secrecy if the probability $p_{nf}$ is such that $p_{nf}=\calO(\frac{1}{n^\alpha})$ for some $\alpha>1$.
\end{lemma}
\begin{IEEEproof}
We can lower bound $\avgH{\rvM|\rvZ^n}$ as
\begin{align*}
  \avgH{\rvM|\rvZ^n} &\geq \avgH{\rvM|\rvZ^n, \mathrm{rank}(G_\mu)} \\
	& \geq \avgH{\rvM|\rvZ^n, G_\mu\textrm{ is full rank}}\P{G_\mu\textrm{ is full rank}}\\
  &= k (1-p_{nf}) = k - R_sn{p_{nf}}
\end{align*}
If $p_{nf}=\calO(\frac{1}{n^\alpha})$, then $\avgI{\rvM;\rvZ^n} = k -\avgH{\rvM|\rvZ^n}\leq \calO(\frac{1}{n^{\alpha-1}})$, which can be made arbitrary small for $n$ sufficiently large and $\alpha>1$.
\end{IEEEproof}

Let $C^n(\lambda,\rho)$ be an LDPC ensemble with $n$ variable nodes, left edge degree distributions $\lambda(x)=\sum_{i\geq 1}\lambda_ix^{i-1}$ and right node degree distribution $\rho(x)=\sum_{i\geq 1}\rho_i x^{i-1}$~\cite[\S 3.4]{ModernCodingTheory} with possibly some expurgations. The degree distributions $\lambda(x), \rho(x)$ are from an edge perspective, that is $\lambda_i$ is the fraction of edges connected to a variable node of degree $i$ and $\rho_j$ is similarly defined.

Let $P_e^{(n)}(\epsilon)$ denote the probability of block error for codes from $C^n(\lambda, \rho)$ over BEC$(\epsilon)$ under iterative decoding. An important interpretation of $P_e^{(n)}(\epsilon)$ is the following: for a parity-check matrix $H$ with degree distribution $(\lambda,\rho)$, $1-P_e^{(n)}(\epsilon)$ is a lower bound on the probability that erased columns of $H$ (over a BEC$(\epsilon)$) form a full-rank submatrix. Using this interpretation and results from \cite{Thangaraj2007}, we have the following immediate corollary of Lemma~\ref{lm:secrecy_perr_connection}.

\begin{corollary}
\label{cor:secrecy_block_error}
If there exists $\epsilon^* > 0$ such that $P_e^{(n)}(\epsilon) = \calO(\frac{1}{n^\alpha})$,  ($\alpha > 1$) for $\epsilon < \epsilon^*$, then the dual of a code from $C^n(\lambda, \rho)$ used in a coset coding scheme provides strong secrecy over a $\mathrm{BEWC}(\epsilon)$ for $\epsilon > 1 - \epsilon^*$. 
\end{corollary}

\begin{figure}%
\includegraphics[width=\columnwidth]{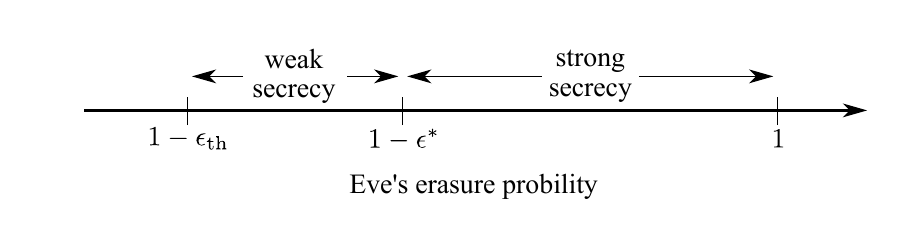}%
\caption{Weak and strong secrecy regions using duals of LDPC codes}%
\label{fig:weak_strong_regions}%
\end{figure}

It is immediately clear that we will have $\epsilon^* \leq \epsilon_\mathrm{th}$, where $\epsilon_\mathrm{th}$ is the erasure threshold for the ensemble over LDPC codes \cite{ModernCodingTheory}. As noted in \cite{Thangaraj2007}, when $\epsilon \leq \epsilon_\mathrm{th}$ we have weak secrecy. In view of this, we will have \emph{guaranteed} weak and strong secrecy regions as illustrated in Fig.~\ref{fig:weak_strong_regions} by doing ``coset coding'' using duals of LDPC codes. We know that degree distributions can be optimized so that $1 - \epsilon_\mathrm{th}$ is very close to the code rate. Since LDPC codes achieve capacity over a BEC, our coding scheme will achieve weak secrecy very close to the secrecy rate and strong secrecy slightly away from the secrecy rate. In the next section, we will show that $\epsilon^*$ exists for some restricted ensembles of LDPC codes.

\section{The LDPC ensemble without short cycles}
\label{sec:ldpc-nosmallcycles}
In this section, we study the sub-ensemble of Tanner graphs \cite{ModernCodingTheory} whose girth is at least $2k$ for some integer $k \geq 2$ which does not change with the block length $n$. We denote the ensemble of all Tanner graphs by $\calG(n, \lambda, \rho)$ and the sub-ensemble of girth $\geq g$ graphs by $\calG_g(n, \lambda, \rho)$. We associate $i$ \emph{sockets} to each node of degree $i$. An edge in a Tanner graph is an unordered pair containing one bit node socket and one check node socket. A Tanner graph with $|E|$ edges has $|E|$ sockets on each side. Therefore, the size of the ensemble equal to the number of permutation of the check node sockets, which is $|E|!$. First we show that the size of our sub-ensemble is not negligible compared to the size of the original ensemble as $n \rightarrow \infty$.

\begin{lemma}[{\cite[Corollary 4]{McKay2004}}]\label{McKay_lemma}
Let $n, g$ be even positive integers and $d \geq 3 $ be an integer. As $n$ grows, let $(d-1)^{2g - 1} = o(n)$. Then, the number of (labeled) $d$-regular bipartite graphs on $n$ vertices with girth greater than $g$ is 
\[
	\frac{(nd/2)!}{(d!)^n} \exp\left(-\sum_{s = 1}^{g/2} \frac{(d-1)^{2s}}{2s} + o(1) \right)
\]
as $n \rightarrow \infty$.
\end{lemma}

Note that the number of $d$-regular bipartite graphs on $n$ vertices is $(nd/2)!/(d!)^n$. The following corollary is then immediate.

\begin{corollary}
Let $g, n$ be positive even numbers and let $d \geq 3$ be an integer. Let $d, g$ remain constant as $n \rightarrow \infty$. Then, the fraction of $(d, d)$ regular bipartite graphs that have girth greater than $g$ is
\[
	\exp\left(-\sum_{s = 1}^{g/2} \frac{(d-1)^{2s}}{2s} + o(1) \right)
\]
as $n \rightarrow \infty$. In particular, this fraction is bounded away from zero for large $n$.
\end{corollary}

\begin{lemma} \label{lemma:girth_fraction}
Let a $(\lambda, \rho)$ irregular Tanner graph ensemble be such that $\max\{\deg(\lambda), \deg(\rho)\} > 2$ and the coefficients of the degree distribution polynomials are rational. Let $g > 0$ be an integer that remains constant with block length $n$. There exists an increasing sequence $(n_k)$ of positive integers such that the fraction of graphs of girth $ > g$ in $\calG(n_k, \lambda, \rho)$ is bounded away from zero as $k \rightarrow \infty$.
\end{lemma}

\begin{IEEEproof}
Let $d$ be the least common multiple of all the vertex degrees in the graph. Clearly, $d > 2$ and it is a function of only $\lambda$ and $\rho$. Let $a$ be the smallest positive integer such that 
\[
	\tfrac{a\tilde{\lambda}_i}{d}, \qquad \tfrac{a\tilde{\rho}_j}{d} \in \mathbb{N}
\]
where $\tilde{\lambda}_i$ is the fraction of variable nodes of degree $i$ and $\tilde{\rho}_j$ is the fraction of check nodes of degree $j$ \cite[\S 3.4]{ModernCodingTheory}. Consider the Tanner graph ensemble with $n_k = ak$ variable nodes. We can group $d/i$ of the degree $i$ variable nodes to get one variable node of degree $d$. If we do this for all the variable node degrees, we will have a left regular Tanner graph with left degree $d$. Similarly, we can repeat this process for the check nodes to get a $(d, d)$ regular Tanner graph. Note that in this node grouping process, we preserve the number of edges since the ensemble allows the possibility of multiple edges. The girth of the resultant regular graph is not more than that of the original graph. It can also be noted that there is a one-one correspondence between the graphs in the $(\lambda, \rho)$ ensemble and those in the $(d, d)$ ensemble. By lemma \ref{McKay_lemma}, the fraction of graphs with girth $> g$ in the $(d, d)$ ensemble, say $\mu$, is non-zero if $k$ is large enough. So, the fraction of graphs in the $(\lambda, \rho)$ ensemble with girth  $> g$ is at least $\mu$. This proves the lemma.
\end{IEEEproof}

\begin{remark}
Let $X$ be a graph dependent positive number. Let $\mathbb{E} X$ represent the expectation of $X$ over $\calG(n, \lambda, \rho)$. Let $\mathbb{E}_1 X$ be the expectation over $\calG_g(n, \lambda, \rho)$ and $\mathbb{E}_2 X$ be the expectation over $\calG(n, \lambda, \rho)\setminus\calG_g(n, \lambda, \rho)$. We have 
\begin{align*}
	\mathbb{E}X = q_n \mathbb{E}_1 X + (1-q_n) \mathbb{E}_2 X
\end{align*}
where $q_n \triangleq |\calG_g(n, \lambda, \rho)| / |\calG(n, \lambda, \rho)|$. By lemma \ref{lemma:girth_fraction}, there exists a $p > 0$ such that for large $n$, we have $q_n \geq p$. Therefore, 
\begin{align*}
	\mathbb{E}X &\geq p \mathbb{E}_1 X \\
	\mathbb{E}_1 X &\leq \frac{1}{p}\mathbb{E} X
\end{align*}
This inequality is used to upper bound $\mathbb{E}_1 X$ when it is easier to find an upper bound to $\mathbb{E} X$.
\end{remark}

\subsection{Stopping sets and stopping number}
For the sake of clarity and completeness, we restate some of the definitions that were originally stated in \cite{Orlitsky2005}. Given a Tanner graph $G$, let $U$ be any subset of variable nodes in $G$. Let the (check node) neighbours of $U$ be denoted by $N(U)$. $U$ is called a \emph{stopping set} if the degree of all the check nodes in the induced subgraph $G[U \cup N(U)]$ is at least two. The \emph{stopping number} of a Tanner graph is defined as the size of its smallest stopping set. For a given Tanner graph, its stopping number is denoted by $s^*$ and the set of all stopping sets is denoted by $\mathbb{S}$. The \emph{stopping ratio} is defined as the ratio of the stopping number to the block length. 

The \emph{average stopping set distribution} is defined as
\[
	E(s) = \mathbb{E}(|\{S \in \mathbb{S} : |S| = s \}|)
\]
where the average is taken over all the Tanner graphs in $\calG(n, \rho, \lambda)$. For any rational $\alpha \in [0, 1]$, it is assumed that there exists a sequence $(n_k)$ of strictly increasing block lengths such that $E(\alpha n_k) > 0$ for all $n_k$. We can then define the \emph{normalized stopping set distribution} as
\[
	\gamma(\alpha) = \lim_{k \rightarrow \infty} \frac{1}{n_k}\log E(\alpha n_k)
\]
It was shown that $\gamma(\alpha)$ is continuous over the set of rationals and hence, it can be extended to a continuous function over $[0, 1]$. The \emph{critical exponent stopping ratio} of a Tanner graph ensemble is defined as
\[
\alpha^* = \inf \{ \alpha > 0 : \gamma(\alpha) \geq 0 \}
\]

\subsection{Block error probability of short-cycle-free ensembles}
In this section, we prove a key result about the average block error probability of short-cycle-free LDPC ensembles, which is central to our claim that the duals of these codes provide strong secrecy. Let $P_B^\mathrm{IT}(C, \epsilon)$ be the probability of block error when the code $C$ is transmitted over $\mathrm{BEC}(\epsilon)$ and iteratively decoded. We define \cite{Orlitsky2005}
\[
	\epsilon_\mathrm{ef} \triangleq \sup\left\{ \epsilon : \max_{\alpha \in [0, \epsilon]} \left( \gamma(\alpha) + (1 - \alpha) h( \tfrac{\epsilon - \alpha}{1 - \alpha} ) - h(\epsilon) \right) \leq 0\right\}
\]
where $h(x)$ is the binary entropy function calculated using natural logarithms. Note that $\gamma(\alpha)$, and $\epsilon_\mathrm{ef}$ are calculated over the entire ensemble $\calG(n, \lambda, \rho)$ instead of the girth-restricted ensemble. Instead of calculating $P_B^\mathrm{IT}(C, \epsilon)$ directly, we take averages of this quantity over an ensemble of codes and show that the average block error probability over the ensemble $\calG_{2k}(n, \lambda, \rho)$ decays as fast as we want it to for $\epsilon < \epsilon_\mathrm{ef}$.
\begin{theorem} \label{thm:block_error}
For $\calG_{2k}(n, \lambda, \rho)$, with minimum variable node degree $l_\mathrm{min}$, maximum variable node degree $l_\mathrm{max}$ and maximum check node degree $r_\mathrm{max} > 2$ we have 
\[
\mathbb{E}_1(P_B^{\mathrm{IT}} (C, \epsilon) ) = \calO\left( \frac{1}{n^{\lceil\frac{ l_\mathrm{min}}{2}k\rceil - k}}\right)
\]
and in the limits of small $\epsilon$ and large $n$
\[
\mathbb{E}_1(P_B^{\mathrm{IT}} (C, \epsilon) ) = \calO\left( \frac{\epsilon^{k}}{n^{\lceil\frac{ l_\mathrm{min}}{2}k\rceil - k}}\right)
\]
\end{theorem}
\begin{IEEEproof}
Let $V_e$ be the set of variable nodes corresponding to the random erasures in the LDPC codeword. The iterative decoding fails iff $V_e$ contains a stopping set. So,
\begin{align*}
P_B^{\mathrm{IT}} (C, \epsilon) = \mathbb{P}(\exists S \in \mathbb{S} : S \subset V_e)
\end{align*}
For any $\delta_1, \delta_2 > 0$, we bound $P_B^{\mathrm{IT}} (C, \epsilon)$ using union bound as
\begin{align*}
P_B^{\mathrm{IT}} (C, \epsilon) & \leq \sum_{i = k}^{\delta_1 n - 1} \left| \left\{ S \in \mathbb{S} : |S| = i \right\}\right| \epsilon^i \\
			& \quad + \mathbb{P}(\exists S \in \mathbb{S} : S \subset V_e, \delta_1 n \leq |S| \leq (\epsilon + \delta_2) n ) \\
			& \quad + \mathbb{P}(\exists S \in \mathbb{S} : S \subset V_e, (\epsilon + \delta_2) n \leq |S| \leq n )
\end{align*}
Using an argument almost identical to the one used in \cite[Theorem 16]{Orlitsky2005}, we can show that the expectations of the second and the third terms go to zero exponentially as $n \rightarrow \infty$ if $\epsilon < \epsilon_\textrm{ef}$. Now,
\begin{align*}
&\mathbb{E}_1\left( \sum_{i = k}^{\delta_1 n - 1} \left| \left\{ S \in \mathbb{S} : |S| = i \right\}\right| \epsilon^i \right) \\
		&\qquad = \sum_{i = k}^{\delta_1 n - 1} \mathbb{E}_1\left( \left| \left\{ S \in \mathbb{S} : |S| = i \right\}\right|\right) \epsilon^i \\
		&\qquad \leq \frac{1}{p}\sum_{i = k}^{\delta_1 n - 1} \mathbb{E}\left( \left| \left\{ S \in \mathbb{S} : |S| = i \right\}\right|\right) \epsilon^i
\end{align*}

A stopping set of $i$ variable nodes can have nodes of different degrees. Let $\calS_i$ denote the set of all non-negative integer solutions to the equation $i_{ l_\mathrm{min}} + i_{ l_\mathrm{min}+1} + \cdots + i_{l_\mathrm{max}} = i$. We can write
\begin{align*}
&\mathbb{E}\left( \left| \left\{ S \in \mathbb{S} : |S| = i \right\}\right|\right) \epsilon^i \\
& \qquad = \epsilon^i \sum_{\{i_s\} \in \calS_i} \tbinom{n\tilde{\lambda}_{ l_\mathrm{min}}}{i_{ l_\mathrm{min}}} \tbinom{n\tilde{\lambda}_{ l_\mathrm{min}+1}}{i_{ l_\mathrm{min}+1}} \cdots \tbinom{n\tilde{\lambda}_{l_\mathrm{max}}}{i_{{l_\mathrm{max}}}} \frac{A}{\binom{|E|}{\sum s i_s}} \\
& \qquad \leq \epsilon^i \binom{n}{i}\sum_{\{i_s\} \in \calS_i} \frac{A}{\binom{|E|}{\sum s i_s}}
\end{align*}
Here, $A$ is the number of ways to connect the selected $i$ variable nodes to form a stopping set. This number is independent of $n$ as long as $i$ is just a small fraction of it. We also note that if we increase the degree of all the check nodes in the graph, $A$ can only increase. Therefore, we may upper bound $A$ by the number of ways to form a stopping set assuming each check node has the maximum possible degree, $r_\mathrm{max}$. The latter number is equal to $\mathrm{coef}\left( \left( (1+x)^{r_\mathrm{max}} - r_\mathrm{max}x \right)^m, x^{\sum s i_s} \right)$ by elementary combinatorics. We have,
\begin{align*}
A & \leq \mathrm{coef}\left( \left( (1+x)^{r_\mathrm{max}} - r_\mathrm{max}x \right)^m, x^{\sum s i_s} \right) \\
	& \leq \binom{m + \lfloor \frac{\sum s i_s}{2} \rfloor - \lceil \frac{\sum s i_s}{r_\mathrm{max}} \rceil}{\lfloor \frac{\sum s i_s}{2} \rfloor} (2r_\mathrm{max}-3)^{\sum s i_s}
\end{align*}
where the last inequality follows from \cite[Lemma 18]{Orlitsky2005}. If we denote $\sum s i_s$ by $w$, we have $i  l_\mathrm{min}\leq w \leq il_\mathrm{max}$. So,
\begin{align*}
&\mathbb{E}\left( \left| \left\{ S \in \mathbb{S} : |S| = i \right\}\right|\right) \epsilon^i \\
				& \qquad \leq \epsilon^i \binom{n}{i}\sum_{\{i_s\} \in \calS_i} \binom{m + \lfloor \frac{w}{2}\rfloor - \lceil \frac{w}{r_\mathrm{max}}\rceil}{\lfloor \frac{w}{2}\rfloor}\frac{(2r_\mathrm{max}-3)^w}{\binom{|E|}{w}} \\
				& \qquad \leq \epsilon^i \binom{n}{i} (2r_\mathrm{max}-3)^{il_\mathrm{max}} \sum_{\{i_s\} \in \calS_i} \binom{m + \frac{il_\mathrm{max}}{2}}{\lfloor \frac{w}{2}\rfloor}\frac{1}{\binom{|E|}{w}} \\
				& \qquad \leq \epsilon^i \binom{n}{i} (2r_\mathrm{max}-3)^{il_\mathrm{max}} \sum_{\{i_s\} \in \calS_i} \frac{\left(m + \frac{il_\mathrm{max}}{2} \right)^{\lfloor \frac{w}{2} \rfloor} w!}{\lfloor \frac{w}{2}\rfloor ! \left( |E| - il_\mathrm{max} \right)^w } 
\end{align*}
If we denote the summand by $f(w)$, we have
\begin{align*}
	\tfrac{f(2r+1)}{f(2r)} &= \tfrac{2r+1}{|E| - il_\mathrm{max}} \leq \tfrac{il_\mathrm{max}}{|E| - il_\mathrm{max}} \leq \
															\tfrac{\delta n l_\mathrm{max}}{|E| - \delta_1 n l_\mathrm{max}} \leq 1
\end{align*}
if we choose $\delta_1$ small enough. Also,
\[
\tfrac{f(2r+2)}{f(2r+1)} = 2 \tfrac{m + \tfrac{il_\mathrm{max}}{2}}{|E| - il_\mathrm{max}} \leq 2 \tfrac{m + \tfrac{\delta_1 n l_\mathrm{max}}{2}}{|E| - \delta_1 n l_\mathrm{max}}
\]
Since $r_\mathrm{max} > 2$ we have $|E| > 2m$. Again, if we choose $\delta_1$ small enough, we will have $f(2r+2)/f(2r+1) \leq 1$. So, $f(w)$ is a non-increasing function and $w \geq i l_\mathrm{min}$. We now have
\begin{align*}
&\mathbb{E}\left( \left| \left\{ S \in \mathbb{S} : |S| = i \right\}\right|\right) \epsilon^i \\
				& \qquad \leq \epsilon^i \tbinom{n}{i} (2r_\mathrm{max}-3)^{il_\mathrm{max}} \\
						& \qquad \qquad \times \sum_{\{i_s\} \in \calS_i} \frac{\left(m + \frac{il_\mathrm{max}}{2} \right)^{\lfloor \frac{i  l_\mathrm{min}}{2} \rfloor} (i  l_\mathrm{min})!}{\lfloor \frac{i  l_\mathrm{min}}{2}\rfloor ! \left( |E| - il_\mathrm{max} \right)^{i  l_\mathrm{min}} } \\
				& \qquad \leq \epsilon^i \tbinom{n}{i} (2r_\mathrm{max}-3)^{il_\mathrm{max}} (i+1)^{r_\mathrm{max}} \\
						&	\qquad \qquad \times \frac{\left(m + \frac{\delta_1 n l_\mathrm{max}}{2} \right)^{\lfloor \frac{i  l_\mathrm{min}}{2} \rfloor} (i  l_\mathrm{min})!}{\lfloor \frac{i  l_\mathrm{min}}{2}\rfloor ! \left( |E| - \delta_1 n l_\mathrm{max} \right)^{i  l_\mathrm{min}} } \\
				& \qquad \leq \epsilon^i \tbinom{n}{i} (2r_\mathrm{max}-3)^{il_\mathrm{max}} \frac{(i+1)^{r_\mathrm{max}}}{n^{\lceil \frac{i  l_\mathrm{min}}{2} \rceil}} \\
						&\qquad \qquad \times \frac{\left(r_0 + \frac{\delta_1 r_\mathrm{max}}{2} \right)^{\lfloor \frac{i  l_\mathrm{min}}{2} \rfloor} (i  l_\mathrm{min})!}{\lfloor \frac{i  l_\mathrm{min}}{2}\rfloor ! \left( r_1 - \delta_1 r_\mathrm{max} \right)^{i  l_\mathrm{min}} } \\
				& \qquad \triangleq \epsilon^i J_i
\end{align*}
Here, $r_0 = m/n$ and $r_1 = |E|/n$ depend only on $\rho$ and $\lambda$. If $i$ remains a constant as $n \rightarrow \infty$, we have
\begin{equation}
J_i = \Theta\left( \frac{1}{n^{\lceil \frac{i  l_\mathrm{min}}{2} \rceil - i}}\right)
\end{equation}
Also,
\begin{align*}
\frac{J_{i+2}}{J_i} &= \frac{\binom{n}{i+2}}{\binom{n}{i}} (2r_\mathrm{max}-3)^{2l_\mathrm{max}} \frac{(r_0 + \frac{\delta_1 l_\mathrm{max}}{2})^{ l_\mathrm{min}}}{(r_1 - \delta_1 l_\mathrm{max})^{2  l_\mathrm{min}}} \\
				& \qquad \times \left(\tfrac{i+3}{i+1}\right)^{r_\mathrm{max}}\frac{(i  l_\mathrm{min} + 2 l_\mathrm{min})! \lfloor \frac{i l_\mathrm{min}}{2}\rfloor!}{(i  l_\mathrm{min})! \left( \lfloor \frac{i  l_\mathrm{min}}{2} \rfloor +  l_\mathrm{min} \right)! n^{ l_\mathrm{min}}} \\
				&\leq \frac{(n-i-1)(n-i)}{(i+1)(i+2)} (2r_\mathrm{max}-3)^{2l_\mathrm{max}} \left(\tfrac{i+3}{i+1}\right)^{r_\mathrm{max}} \\
				& \qquad \times \frac{(r_0 + \frac{\delta_1 l_\mathrm{max}}{2})^{ l_\mathrm{min}}}{(r_1 - \delta_1 l_\mathrm{max})^{2  l_\mathrm{min}}} \
				\frac{(i  l_\mathrm{min} + 2 l_\mathrm{min})^{2  l_\mathrm{min}}}{\left( \lfloor \frac{i  l_\mathrm{min}}{2} \rfloor + 1\right)^{ l_\mathrm{min}} n^{ l_\mathrm{min}}}
\end{align*}
Using $\frac{i+3}{i+1} \leq 2$, $i  l_\mathrm{min} + 2  l_\mathrm{min} \leq 3 i l_\mathrm{min}$, $\lfloor x \rfloor + 1 \geq x$,
\begin{align*}
\frac{J_{i+2}}{J_i} &\leq \frac{n^2}{i^2} (2r_\mathrm{max}-3)^{2l_\mathrm{max}} 2^{r_\mathrm{max}} \frac{(r_0 + \frac{\delta_1 l_\mathrm{max}}{2})^{ l_\mathrm{min}}}{(r_1 - \delta_1 l_\mathrm{max})^{2  l_\mathrm{min}}} \\
				&\qquad \times \frac{(3 i  l_\mathrm{min})^{2  l_\mathrm{min}}}{\left( \frac{i  l_\mathrm{min}}{2} \right)^{ l_\mathrm{min}} n^{ l_\mathrm{min}}}
\end{align*}
Choosing $\delta_3 \in (0, 1)$ such that $r_1 - \delta_3 l_\mathrm{max} > 0$ and $\delta_1 < \delta_3$, 
\begin{align*}
\frac{J_{i+2}}{J_i} &\leq (2r_\mathrm{max}-3)^{2l_\mathrm{max}} 2^{r_\mathrm{max}} \\
										& \qquad \times \frac{(r_0 + \frac{\delta_3 l_\mathrm{max}}{2})^{ l_\mathrm{min}} (3 l_\mathrm{min})^{2 l_\mathrm{min}}}{(r_1 - \delta_3 l_\mathrm{max})^{2  l_\mathrm{min}}\left(\frac{ l_\mathrm{min}}{2}\right)^{ l_\mathrm{min}}}\left(\frac{i}{n}\right)^{ l_\mathrm{min} - 2} \\
				& = B \left(\frac{i}{n}\right)^{ l_\mathrm{min} - 2} \quad \leq \quad B \delta_1^{ l_\mathrm{min} - 2}
\end{align*}
where $B$ depends only on $\lambda$ and $\rho$.
\begin{align*}
&\mathbb{E}_1\left( \sum_{i = k}^{\delta_1 n - 1} \left| \left\{ S \in \mathbb{S} : |S| = i \right\}\right| \epsilon^i \right) \leq \frac{1}{p} \sum_{i = k}^{\delta_1 n - 1} \epsilon^i J_i \\
& \qquad \leq \frac{1}{p} \epsilon^{k} \sum_{i = k}^{\delta_1 n - 1} J_i \\
& \qquad = \frac{1}{p} \epsilon^{k} \left[ \Theta\left( \frac{1}{n^{\lceil \frac{ l_\mathrm{min}}{2}k\rceil - k}}\right) + \Theta\left( \frac{1}{n^{\lceil \frac{ l_\mathrm{min}}{2}(k+1)\rceil - k-1}}\right)\right] \\
& \qquad \qquad \times \sum_{i = 0}^{\delta_1 n/2} \left( B \delta_1^{ l_\mathrm{min} - 2}\right) ^i
\end{align*}
If $\delta_1$ is small enough, then the summation in the above equation is bounded by a decreasing geometric sum. So,
\begin{align}
\mathbb{E}_1\left( \sum_{i = k}^{\delta_1 n - 1} \left| \left\{ S \in \mathbb{S} : |S| = i \right\}\right| \epsilon^i \right) &= \calO\left( \frac{\epsilon^{k}}{n^{\lceil \frac{ l_\mathrm{min}}{2}k\rceil - k}}\right) \notag \\
\Rightarrow \mathbb{E}_1 \left( P_B^\textrm{IT} (C, \epsilon) \right) &= \calO\left( \frac{\epsilon^{k}}{n^{\lceil \frac{ l_\mathrm{min}}{2}k\rceil - k}}\right)
\end{align}
as $\epsilon \rightarrow 0$ and $n \rightarrow \infty$.
\end{IEEEproof}
From the above theorem, the average block error probability in our ensemble decays faster than $\frac{1}{n^2}$ for $l_\mathrm{min} > 2$ and $k > 3$. This correpsonds to LDPC ensembles with a minimum bit node degree of at least 3 and girth at least 4. By corollary \ref{cor:secrecy_block_error}, the duals of these LDPC codes achieve strong secrecy over a BEWC of erasure probability $1 - \epsilon_\mathrm{ef}$. 

The (3, 6) regular LDPC ensemble has $\epsilon_\mathrm{th} = 0.429$, $\epsilon_\mathrm{ef} = 0.366$ and rate $1/2$. When duals of codes in this ensemble are used on $\mathrm{BEWC}(\epsilon)$, a secret communication rate of 0.5 is achieved with weak secrecy when $\epsilon \in (0.571, 0.634)$ and with strong secrecy when $\epsilon > 0.634$. Our numerical calculations indicate that some of the degree distributions that are optimized for very high $\epsilon_\mathrm{th}$ have $\epsilon_\mathrm{ef} < 0.366$. 

\section{Conclusion and future directions}
\label{sec:conclusion}
In this work, we have shown that duals of LDPC codes with girth greater than 4 and minimum left degree at least $3$ achieve strong secrecy on the binary erasure wiretap channel. LDPC ensembles with degree 2 nodes play an important role in achieving capacity on the binary erasure channel. Further study is required on the relationship between these LDPC codes and strong secrecy. Another research possibility involves optimizing the degree distributions to find LDPC ensembles with a very high $\epsilon_\mathrm{ef}$ for a given rate. 

\bibliographystyle{IEEEtran}
\bibliography{itw2010}

\begin{thebibliography}{10}
\providecommand{\url}[1]{#1}
\csname url@samestyle\endcsname
\providecommand{\newblock}{\relax}
\providecommand{\bibinfo}[2]{#2}
\providecommand{\BIBentrySTDinterwordspacing}{\spaceskip=0pt\relax}
\providecommand{\BIBentryALTinterwordstretchfactor}{4}
\providecommand{\BIBentryALTinterwordspacing}{\spaceskip=\fontdimen2\font plus
\BIBentryALTinterwordstretchfactor\fontdimen3\font minus
  \fontdimen4\font\relax}
\providecommand{\BIBforeignlanguage}[2]{{%
\expandafter\ifx\csname l@#1\endcsname\relax
\typeout{** WARNING: IEEEtran.bst: No hyphenation pattern has been}%
\typeout{** loaded for the language `#1'. Using the pattern for}%
\typeout{** the default language instead.}%
\else
\language=\csname l@#1\endcsname
\fi
#2}}
\providecommand{\BIBdecl}{\relax}
\BIBdecl

\bibitem{Shannon1948a}
C.~E. Shannon, ``{C}ommunication {T}heory of {S}ecrecy {S}ystems,'' \emph{Bell
  System Technical Journal}, vol.~28, pp. 656--715, 1948.

\bibitem{Wyner1975}
A.~D. Wyner, ``{T}he {W}ire-{T}ap {C}hannel,'' \emph{Bell System Technical
  Journal}, vol.~54, no.~8, pp. 1355--1367, October 1975.

\bibitem{Maurer2000}
U.~M. Maurer and S.~Wolf, ``{I}nformation-{T}heoretic {K}ey {A}greement: {F}rom
  {W}eak to {S}trong {S}ecrecy for {F}ree,'' in \emph{Advances in {C}ryptology
  - {E}urocrypt 2000}, Lecture Notes in Computer Science.\hskip 1em plus 0.5em
  minus 0.4em\relax B. Preneel, 2000, p. 351.

\bibitem{Csiszar1996}
I.~Csisz\'ar, ``{A}lmost {I}ndependence and {S}ecrecy {C}apacity,''
  \emph{Problems of Information Transmission}, vol.~32, no.~1, pp. 40--47,
  January-March 1996.

\bibitem{Bennett1995}
C.~H. Bennett, G.~Brassard, C.~Cr{\'e}peau, and U.~Maurer, ``{G}eneralized
  {P}rivacy {A}mplification,'' \emph{IEEE Trans. Inf. Theory}, vol.~41, no.~6,
  pp. 1915--1923, November 1995.

\bibitem{Ozarow1984}
L.~H. Ozarow and A.~D. Wyner, ``{W}ire {T}ap {C}hannel {II},'' \emph{A{T}\&{T}
  {B}ell {L}aboratories {T}echnical {J}ournal}, vol.~63, no.~10, pp.
  2135--2157, December 1984.

\bibitem{Thangaraj2007}
A.~Thangaraj, S.~Dihidar, A.~R. Calderbank, S.~W. McLaughlin, and J.-M.
  Merolla, ``{A}pplications of {LDPC} {C}odes to the {W}iretap {C}hannels,''
  \emph{IEEE Trans. Inf. Theory}, vol.~53, no.~8, pp. 2933--2945, Aug. 2007.

\bibitem{Liu2007}
R.~Liu, Y.~Liang, H.~V. Poor, and P.~Spasojevi\'c, ``{S}ecure {N}ested {C}odes
  for {T}ype {II} {W}iretap {C}hannels,'' in \emph{Proceedings of IEEE
  Information Theory Workshop}, Lake Tahoe, California, USA, September 2007,
  pp. 337--342.

\bibitem{Cohen2006}
G.~Cohen and G.~Zemor, ``{S}yndrome-{C}oding for the {W}iretap {C}hannel
  {R}evisited,'' in \emph{Proc. IEEE Information Theory Workshop}, Chengdu,
  China, October 2006, pp. 33--36.

\bibitem{Di2002}
C.~Di, D.~Proietti, I.~Telatar, T.~Richardson, and R.~Urbanke, ``Finite-length
  analysis of low-density parity-check codes on the binary erasure channel,''
  \emph{Information Theory, IEEE Transactions on}, vol.~48, no.~6, pp. 1570
  --1579, jun 2002.

\bibitem{Richardson2001c}
T.~Richardson and R.~Urbanke, ``The capacity of low-density parity-check codes
  under message-passing decoding,'' \emph{Information Theory, IEEE Transactions
  on}, vol.~47, no.~2, pp. 599 --618, feb 2001.

\bibitem{Orlitsky2005}
A.~Orlitsky, K.~Viswanathan, and J.~Zhang, ``{S}topping set distribution of
  {LDPC} code ensembles,'' \emph{IEEE Transactions on Information Theory},
  vol.~51, no.~3, pp. 929 --953, march 2005.

\bibitem{Olgica}
O.~Milenkovic, E.~Soljanin, and P.~Whiting, ``Asymptotic spectra of trapping
  sets in regular and irregular ldpc code ensembles,'' \emph{Information
  Theory, IEEE Transactions on}, vol.~53, no.~1, pp. 39 --55, jan. 2007.

\bibitem{Burshtein}
D.~Burshtein and G.~Miller, ``Asymptotic enumeration methods for analyzing ldpc
  codes,'' \emph{Information Theory, IEEE Transactions on}, vol.~50, no.~6, pp.
  1115 -- 1131, june 2004.

\bibitem{ModernCodingTheory}
T.~Richardson and R.~Urbanke, \emph{{M}odern {C}oding {T}heory}.\hskip 1em plus
  0.5em minus 0.4em\relax Cambridge University Press, 2008.

\bibitem{Korada}
S.~Korada and R.~Urbanke, ``Exchange of limits: Why iterative decoding works,''
  in \emph{Information Theory, 2008. ISIT 2008. IEEE International Symposium
  on}, july 2008, pp. 285 --289.

\bibitem{Amraoui2009}
A.~Amraoui, A.~Montanari, T.~Richardson, and R.~Urbanke, ``Finite-length
  scaling for iteratively decoded ldpc ensembles,'' \emph{Information Theory,
  IEEE Transactions on}, vol.~55, no.~2, pp. 473 --498, feb. 2009.

\bibitem{McKay2004}
B.~D. McKay, N.~C. Wormald, and B.~Wysocka, ``{S}hort cycles in random regular
  graphs,'' \emph{Electr. J. Comb.}, vol.~11, no.~1, 2004.

\end{thebibliography}
\end{document}